# Dzyaloshinskii–Moriya interaction across antiferromagnet / ferromagnet interface

Xin Ma[1,*,+], Guoqiang Yu[2,*], Seyed A. Razavi[2], Stephen S. Sasaki[3], Xiang Li[2], Kai Hao[1], Sarah H. Tolbert[3],

Kang L. Wang[2,4], and Xiaoqin Li[1,‡]

[1]Department of Physics, Center for Quantum systems, The University of Texas at Austin, Austin, Texas 78712, USA
[2]Department of Electrical Engineering, University of California, Los Angeles, California 90095, USA
[3]Department of Chemistry and Biochemistry, University of California, Los Angeles, California 90095, USA
[4]Department of Physics, University of California, Los Angeles, California 90095, USA

The antiferromagnet (AFM) / ferromagnet (FM) interfaces are of central importance in recently developed pure electric or ultrafast control of FM spins, where the underlying mechanisms remain unresolved. Here we report the direct observation of Dzyaloshinskii–Moriya interaction (DMI) across the AFM/FM interface of IrMn/CoFeB thin films. The interfacial DMI is quantitatively measured from the asymmetric spin-wave dispersion in the FM layer using Brillouin light scattering. The DMI strength is enhanced by a factor of 7 with increasing IrMn layer thickness in the range of 1- 7.5 nm. Our findings provide deeper insight into the coupling at AFM/FM interface and may stimulate new device concepts utilizing chiral spin textures such as magnetic skyrmions in AFM/FM heterostructures.

Control of spins in ferromagnets (FMs) utilizing antiferromagnets (AFMs) is an emerging branch of spintronics[1-5]. By placing an AFM layer adjacent to the FM layer, the unique electric, magnetic and transport properties of the AFM may be used to control the FM layer via interfacial coupling. Conventionally, the AFM layer has mostly played a passive role in device operations by either improving the hardness of FM via exchange bias[6-8] or increasing the magnetic damping of FM through spin pumping[9-13]. More recently, the AFMs have been used as active control elements, leading to promising breakthroughs in the electric and ultrafast control of FM spins. For instance, electric current-induced magnetization switching of FM without an external magnetic field has been realized in the AFM/FM systems[1-4]. These pioneering experiments have been shown to generate the pure spin current in the AFM or at the AFM/FM interface[1, 2, 14-16] and to utilize the exchange bias to break the switching symmetry[1-4]. Moreover, coherent spin precession in the FM layer can be effectively excited by an ultrafast spin-exchange-coupling torque across the AFM/FM interface[5]. The laser pulse perturbs the AFM spin arrangement, which in turn generates an intense and non-thermal transient torque acting on the FM spins. Despite these promising achievements, certain limitations such as the incomplete magnetization switching by current remain in the AFM/FM system. Thus, elucidating interaction mechanisms across the AFM/FM interface is not only important from a scientific point of view, but also of great technologic relevance.

In heterostructures with broken spatial inversion symmetry, the interfacial Dzyaloshinskii–Moriya interaction (DMI) has been identified as an important mechanism leading to a host of interesting phenomena. DMI promotes non-collinear spin alignments and determines the chirality and dynamics of chiral spin textures[17-19]. For instance, DMI stabilizes the magnetic skrymions and domain walls in the Néel configuration with certain chirality and lends a mechanism for driving skrymion and domain wall motion via spin torques[20-25]. Similarly, DMI likely contributes to the current-induced magnetization switching in the AFM/FM systems, because such switching may occur via magnetic domain nucleation followed by spin-torque-driven domain wall propagation[16, 26-28]. However, no direct experimental observation of DMI across the AFM/FM interface has been reported previously.

In this letter, we report quantitative measurements of interfacial DMI in IrMn/CoFeB/MgO multilayer thin films. The DMI coefficient $D$ is obtained from the asymmetric spin wave dispersion in the CoFeB layer probed with Brillouin light scattering (BLS). $D$ is inversely proportional to the CoFeB thickness, indicating the interfacial nature of the observed DMI. On the other hand, the coefficient $D$ continuously increases in magnitude by a factor of 7 when the IrMn layer thickness increases from 1 to 7.5 nm. There are important differences as well as similarities between the DMI in the AFM/FM system reported here and that in heavy metal (HM)/FM bilayers investigated extensively in recent years[29-40]. Our discovery is in synergy with many on-going activities exploring analogous phenomena between HM/FM and AFM/FM bilayers[1, 2, 14-16]. The rich interaction phenomena in the AFM/FM systems may enable effective control of magnetic skyrmions and domain walls.

A series of $Ir_{22}Mn_{78}(t)/Co_{20}Fe_{60}B_{20}(2)/MgO(2)/Ta(2)$ thin films were deposited by magnetron sputtering at room temperature on thermally oxidized silicon substrates, where the subscript represents the percentage of each element in the alloyed layer and the numbers in parentheses denote the nominal layer thicknesses in nanometers. Moreover, Ir(5)/CoFeB(1.2)/MgO/Ta and IrMn(5)/CoFeB(0.8-2, wedge)/ MgO/Ta thin films were prepared under the same conditions. Following the deposition, all multilayer thin films were further annealed at 250 °C for 30 minutes. For the field cooling purpose, an in-plane magnetic field of 6 kOe was applied during the annealing procedure to establish in-plane exchange bias (EB) in the IrMn/CoFeB thin films. The IrMn layer is poly-crystalline and likely exhibits a non-collinear anti-ferromagnetic spin alignment as suggested by spin-orbit torque measurements [14, 16] and neutron diffraction studies[41] on similar samples.

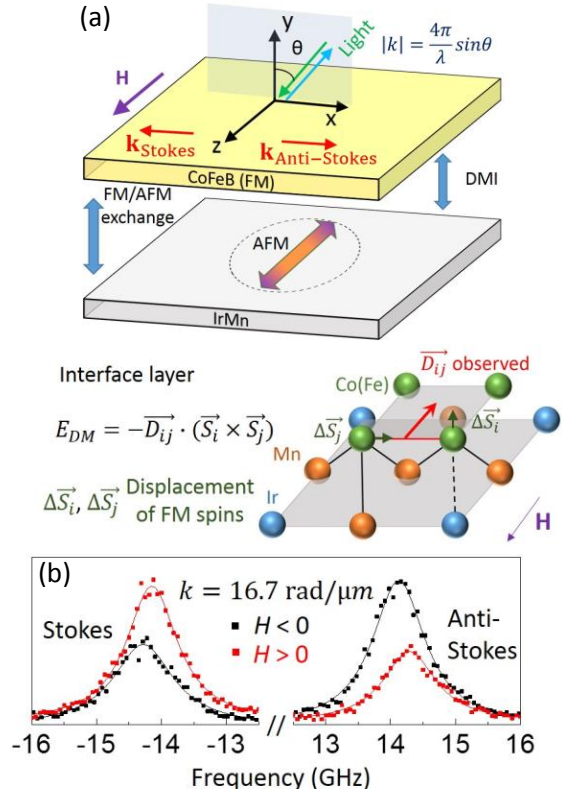

Fig. 1. (a) Schematics of BLS experiment and possible atomic arrangement at the interface. (b) BLS spectra for DE spin waves recorded at a fixed incident angle with $k = 16.7$ rad/μm under oppositely oriented external magnetic fields **H**. The solid lines represent fittings with Lorentzian functions.

We measured the spin wave dispersion in the CoFeB layer using BLS in a geometry shown in Fig. 1a[36]. An in-plane magnetic field **H** was applied along the $z$ axis in all measurements. A laser beam with s-linear polarization was incident on the sample, and the p-polarized component of the backscattered light was collected and sent to a Sandercock-type multipass tandem Fabry-Perot interferometer. In the light scattering process, the total momentum is conserved in the plane of the thin film. As a result, the Stokes (anti-Stokes) peaks in BLS spectra correspond to the creation (annihilation) of magnons with wave vector $|k| = \frac{4\pi}{\lambda}\sin\theta$ along $-x$ ($+x$) direction as illustrated in Fig. 1a, where $\lambda = 532$ nm is the laser wavelength, and $\theta$ refers to the incident angle of light. In order to reduce the uncertainty in $k$, the laser beam was barely focused and an additional spatial filter was placed in the signal collection path. Each BLS spectrum was taken with 17 GHz free spectrum range with 400 channels, and accumulated over 20 minutes. A high signal-to-noise ratio of the measured BLS spectra is critical for accurately determining the measured spin wave frequency. The high quality of CoFeB with Ta seed layer may have contributed to the strong magnon signal[35].

The spin waves probed here are Damon-Eshback (DE) modes with propagation directions perpendicular to **H** (**M**). The spin wave dispersion is described by[30, 31]

$$f = \frac{\gamma}{2\pi}\sqrt{\left(H_{eff} + \frac{2A}{M_s}k^2 + 4\pi M_s(1-\xi(kL)) - \frac{2K_\perp}{M_s}\right) * \left(H_{eff} + \frac{2A}{M_s}k^2 + 4\pi M_s\xi(kL)\right)} - \varepsilon(\mathbf{H}_{EB}, K_\perp, k * sgn(M_z)) - sgn(M_z)\frac{\gamma}{\pi M_s}Dk \quad (1)$$

where $H_{eff} = |\mathbf{H} + \mathbf{H}_{EB}|$ is the magnitude of the effective field by adding the vectors of external field **H** and the equivalent field induced by exchange bias $\mathbf{H}_{EB}$, $\gamma$ is the gyromagnetic ratio, $A$ is the exchange stiffness constant, $\xi(kL) = 1 - (1 - e^{-|kL|})/|kL|$ with $L$ the CoFeB thickness, $K_\perp$ is the interfacial magnetic anisotropy which mainly originates from the CoFeB/MgO interface, $\varepsilon(\mathbf{H}_{EB}, K_\perp, k)$ describes a correction in frequency as discussed below, and $D$ is the DMI coefficient. Both $D$ and $k$ can be positive or negative values in the formula. Detailed justifications of Eq.1 can be found in supplementary information[42]. In Eq.1, the first term on the right hand side describes the spin wave dispersion under mean-field approach and without DMI, which is even in $k$. The second term originates from the non-reciprocity of DE mode spin waves in the presence of interfacial magnetic anisotropy $K_\perp$ and EB. The spin waves propagating along the $-x$ ($+x$) direction, as illustrated in Fig. 1a, localize near the top (bottom) surface of the CoFeB layer. Consequently, the spin waves propagating along $-x$ ($+x$) direction experience a stronger $K_\perp$ (EB), leading to an additional frequency correction as denoted by $\varepsilon(\mathbf{H}_{EB}, K_\perp, k)$. Our experiment and simulation show this second term is much smaller than the DMI effect in samples with a 2.5 nm or thicker InMn layer (see supplementary information)[42]. We take into account this second term explicitly in all analyses of DMI. Most importantly, the third term accounts for the frequency difference between counter-propagating spin waves induced by DMI and is odd in $k$.

Interfacial DMI in the AFM/FM heterostructure is manifested in the lifted chiral degeneracy of the DE spin waves in the CoFeB layer. Figure 1b shows typical BLS spectra for the DE spin waves from the IrMn(5)/CoFeB(2) thin film subject to **H** fields with opposite directions. The most prominent feature is that the frequencies of the Stokes and anti-Stokes peaks (the spin waves with the same $|k|$ but opposite chirality) are different, while such frequency difference changes its sign upon reversing the **H** direction. The asymmetric shift in DE spin wave dispersion is consistent with the frequency shift due to DMI as described by the third term in Eq. 1.

To quantify the DMI coefficient $D$, momentum-resolved

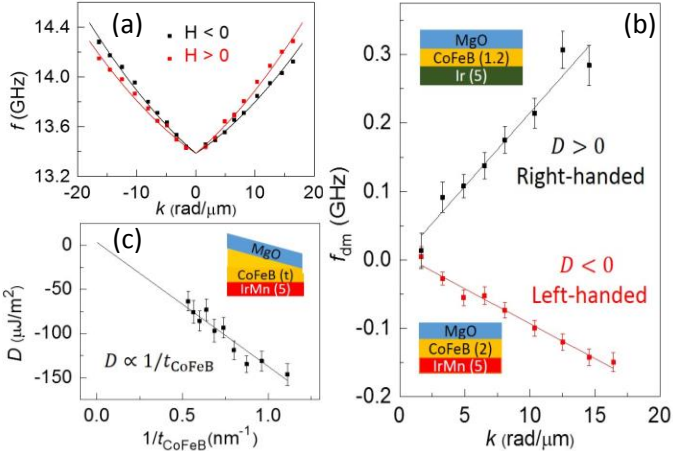

Fig. 2. (a) The asymmetric spin wave dispersion under oppositely oriented **H** at IrMn(5)/CoFeB(2). Solid lines refer to fitting with Eq.1. (b) The linear dependence of $f_{dm}$ on $k$ in IrMn(5)/CoFeB(2) (red) and Ir(5)/CoFeB(1.2) (black) samples. (c) The $D$ as a function of $1/t_{CoFeB}$ at Ir/CoFeB(wedge)/MgO. The solid lines refer to the least square fits.

BLS measurements were performed by varying the light incident angle[30-37, 43]. Figure 2a shows the asymmetric spin wave dispersion at the IrMn(5)/CoFeB film under opposite **H**, which can be well fitted with Eq.1. According to Eq .1, we can simplify the determination of $D$ by subtracting the two spectra.

$$f_{dm} = \frac{\left((f(-k,M_z)-f(k,M_z))-(f(-k,-M_z)-f(k,-M_z))\right)}{2} = \frac{2\gamma}{\pi M_S}Dk + \Delta\varepsilon(k) \quad (2)$$

where $\Delta\varepsilon(k) = \varepsilon(H_{EB}, K_\perp, k) - \varepsilon(H_{EB}, K_\perp, -k)$ is much smaller than the first term $\frac{2\gamma}{\pi M_S}Dk$ in our samples with IrMn thickness $t_{IrMn} \geq 2.5$ nm[42]. This subtraction also removes a possible instrument frequency offset between the Stokes and anti-Stokes peaks. According to Eq. 2, one expects an linear correlation between $f_{dm}$ and $k$, and the slope can be used to determine $D$ after correcting for $\Delta\varepsilon(k)$. The experimental observation in Fig. 2b (red data points and linear fit) is consistent with Eq. 2. The negative slope indicates that $D < 0$ and the left-handed magnetic chirality is preferred in the IrMn(5)/CoFeB(2) film[36].

To rule out the possibility that the observed interfacial DMI could simply arise from Ir atoms, we measured a control sample Ir/CoFeB as shown in Fig. 2b. The signs of the $D$ in Ir/CoFeB (black data points and linear fit) and IrMn/CoFeB (red) are opposite. This clear difference suggests that the interfacial DMI observed in the samples with an IrMn layer is strongly influenced by the Mn atoms (illustrated in Fig. 1a), instead of originating from the contribution of Ir atoms alone. Previous experiments on HM/FM bilayer have reported DMI constants with opposite signs in similar HM/FM bilayers[37-40]. We note that the DMI sign for our Ir/CoFeB/MgO sample is opposite to that measured by Kim *et al*. in Ir/Co/AlOx thin films[37]. Theoretical calculations show that DMI changes sign for Ir/Co and Ir/Fe due to the modification of 3d-5d hybridization near the Fermi level[44]. In our CoFeB alloys, the higher percentage of Fe may have led to the predicted DMI sign change from Ir/Co/AlOx thin films. Different from the previous experiments on HM/FM systems[37-40], the DMI sign change observed here between Ir/CoFeB and IrMn/CoFeB is caused by Mn atoms with AFM spin alignment.

We demonstrate that the measured DMI is an interfacial effect by studying the CoFeB thickness dependence. In many previous studies on magnetic multilayers, the inverse proportionality to the FM thickness is considered as evidence for interfacial effects. Examples include EB[45] and interfacial magnetic anisotropy[46]. Similarly, an inverse proportionality between the $D$ and $t_{CoFeB}$ is observed in the IrMn(5)/CoFeB (wedge) thin film as shown in Fig. 2c.

Next, we show that such interfacial DMI is enhanced by increasing the thickness of IrMn layer. Figure 3a displays the $k$-dependence of $f_{dm}$ in a series of samples where the IrMn underlayer thickness $t_{IrMn}$ is increased from 1-7.5 nm. As $t_{IrMn}$ increases, the slope of the linear fitting increases in magnitude. To investigate the origin of the asymmetric frequency shift of spin waves in the series of samples, we measured the systematic changes of various magnetic parameters for all samples and summarized results in Table 1 including the extracted $D$ values. Notably, the $M_S$ varies only slightly among samples, and the contribution from the second term in Eq.1 remains small[42]. Thus, we conclude that the observed changes in spin wave dispersion originate from the increased $D$ in thin films with thicker IrMn layer.

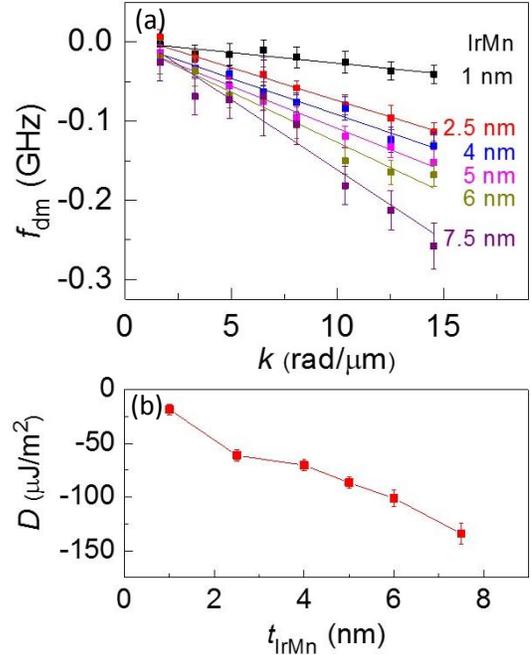

Fig. 3. (a) The linear dependence of $f_{dm}$ on $k$ for a series of IrMn/CoFeB(2) thin films with different IrMn thicknesses. (b) The magnitude of DMI coefficient $D$ increases with larger IrMn thickness. The negative value of $D$ shows that left-handed chirality is favored in this material system.

| IrMn thickness (nm) | 1 | 2.5 | 4 | 5 | 6 | 7.5 |
|---|---|---|---|---|---|---|
| $D(\mu J/m^2)$ [k-BLS] | 18±4 | 61±5 | 70±6 | 86±6 | 101±8 | 134±10 |
| $M_S$ (emu/cm$^3$) [VSM] | 991±50 | 909±45 | 940±47 | 995±50 | 970±48 | 908±45 |
| $2K_\perp/M_S$ (kOe) [H-BLS] | 2.89±0.62 | 2.19±0.55 | 2.42±0.57 | 2.88±0.62 | 2.67±0.58 | 2.17±0.54 |
| $A(10^{-6}$erg/cm) [k-BLS] | 3.17±0.32 | 2.86±0.45 | 2.90±0.34 | 3.25±0.25 | 2.75±0.32 | 2.24±0.43 |
| $\Delta\varepsilon$ (GHz) [Simulation] | -0.0084 | -0.0064 | -0.0070 | -0.0083 | -0.0077 | -0.0063 |
| $H_{EB}$ (Oe) [MOKE] | 0±10 | 0±10 | 10±10 | 25±10 | 77±10 | 350±10 |

Table 1. Magnetic parameters determined on different samples. $\Delta\varepsilon = \varepsilon(H_{EB}, K_\perp, k) - \varepsilon(H_{EB}, K_\perp, -k)$ @ $k = 16.7$ rad/μm

The dependence of DMI on the AFM thickness in AFM/FM is rather different from that on the HM thickness in HM/FM systems. In the HM/FM systems, theoretical and experimental studies on HM thickness dependence suggest that the contribution to DMI is dominated by the spin-orbit coupling of the first atomic layer of HM at HM/FM interface and extends weakly away from the interface[29, 36, 47]. Empirically, the DMI increases with larger $t_{HM}$ but quickly saturates when $t_{HM}$ approaches the spin diffusion length in the HM (e.g. ~2 nm for Pt)[29]. The situation in an AFM/FM heterostructure, however, is more complicated. As observed here, the DMI keeps increasing with a thicker IrMn layer even to the thickness range where $t_{IrMn}$ is approximately one order of magnitude larger than IrMn's spin diffusion length (~ 0.7 nm)[48].

We speculate that the surprising enhancement of $D$ with increasing IrMn layer thickness beyond the IrMn's spin diffusion length is correlated with the AFM spin arrangement of IrMn. In a thicker IrMn layer, thermal fluctuations of the AFM spin arrangement have been suppressed as suggested by other types of experiments on ultrathin IrMn/FM films. For instance, it has been experimentally demonstrated that the AFM grain size increases and the number of unstable grain is reduced in a thicker IrMn layer, leading to an enhanced thermal stability of the AFM order in the IrMn thin film[6,49]. Moreover, less fluctuations of AFM spin arrangement in thicker IrMn layer are suggested by the increase of magnetic order transition temperature via spin pumping experiments in NiFe/Cu/IrMn thin films[9], and by EB and coercivity measurements in IrMn/FM thin films[6, 50]. In view of the experimental challenges to directly probe AFM spin arrangement in nm-thick IrMn layers and quantify their fluctuations, further theoretical studies are necessary to articulate the relation between AFM spin arrangement and the observed DMI.

One substantial benefit in utilizing an AFM layer instead of a HM layer in the multilayer structures is to replace the external magnetic field application with EB, which has led to many technology advancements[1, 2]. Thus, we investigated the possibility of establishing EB in the same films where DMI is observed. We performed Magneto-optical Kerr effect (MOKE) experiments to measure the EB values through in-plane magnetic hysteresis loops for the IrMn($t_{IrMn}$)/CoFeB(2) samples with different $t_{IrMn}$ (data included in supplementary information)[42]. The values of EB are summarized in Table 1. The EB is only clearly established in samples with $t_{IrMn} \geq 4\ nm$. The observation of an enhanced EB in samples with thicker $t_{IrMn}$ layer is consistent with other reports[6, 50].

The observed increase of DMI and EB with thicker IrMn layers in our experiments should not be interpreted as a causal relation between DMI and EB as suggested by recent theoretical studies[41, 42]. One clear evidence is that DMI remains almost unchanged between IrMn/CoFeB samples with and without EB (i.e. with and without field cooling, see supplementary information[42]). Although both DMI and EB are related to the AFM spin arrangement, EB originates from the pinned uncompensated spins of IrMn which is only 4~6% of the interfacial AFM spins[51]. This lack of strong correlation between DMI and EB offers an opportunity to optimize these parameters somewhat independently for device applications.

In conclusion, we directly observed and quantitatively evaluated interfacial DMI in IrMn/CoFeB/MgO multilayer thin films. The DMI is enhanced by a factor of 7 by increasing the IrMn thickness well beyond the spin diffusion length, overcoming a bottleneck for improving DMI via increasing the HM layer thickness in the HM/FM bilayers. We suggest that the enhanced $D$ in a thicker IrMn film originates from less fluctuations of AFM spin arrangement in the IrMn layer suggested by other experiments. The microscopic origin of DMI in the AFM/FM system is likely different from that in the HM/FM systems. Our finding may help interpret the incomplete switching of magnetization driven by electric current in the IrMn(PtMn)/FM systems[1, 2], where an enhanced DMI with a thicker AFM layer raises the threshold of EB induced field required for a complete swiching[26]. To explore AFM/FM systems for engineering chiral spin textures, the magnitude of DMI needs to be further increased by exploring alternative AFM/FM materials. By adding interfacial DMI as a control parameter, a judicious optimization of a number of coupling mechanisms in AFM/FM systems (e.g. DMI, EB and spin torques) may enable improved spintronic devices.

* X. M. and G. Y. contributed equally to this work.
+Email address: xma518@utexas.edu
‡Email address: elaineli@physics.utexas.edu

## ACKNOWLEDGEMENTS

We acknowledge Xiang Li and Jianshi Zhou for trainings on VSM experiment and the usage of the VSM instrument. We also acknowledge Yizheng Wu, Hua Chen, Pantelis Lapas, and Allan H. Macdonald for insightful discussion. The collaboration between UT-Austin and UCLA are supported by SHINES, an Energy Frontier Research Center funded by the U.S. Department of Energy (DoE), Office of Science, Basic Energy Science (BES) under award # DE-SC0012670.

## Reference

1. S. Fukami, C. Zhang, S. DuttaGupta, A. Kurenkov, and H. Ohno, Nat Mater **15**, 535 (2016).

2. Y.-W. Oh, S.-h. Chris Baek, Y. M. Kim, H. Y. Lee, K.-D. Lee, C.-G. Yang, E.-S. Park, K.-S. Lee, K.-W. Kim, G. Go, J.-R. Jeong, B.-C. Min, H.-W. Lee, K.-J. Lee, and B.-G. Park, Nat Nano **11**, 878 (2016).

3. Y.-C. Lau, D. Betto, K. Rode, J. M. D. Coey, and P. Stamenov, Nat Nano **11**, 758 (2016).

4. A. van den Brink, G. Vermijs, A. Solignac, J. Koo, J. T. Kohlhepp, H. J. M. Swagten, and B. Koopmans, Nature Communications **7**, 10854 (2016).

5. X. Ma, F. Fang, Q. Li, J. Zhu, Y. Yang, Y. Z. Wu, H. B. Zhao, and G. Lüpke, Nature Communications **6**, 8800 (2015).

6. K. O'Grady, L. E. Fernandez-Outon, and G. Vallejo-Fernandez, Journal of Magnetism and Magnetic Materials **322**, 883 (2010).

7. J. Nogués and I. K. Schuller, Journal of Magnetism and Magnetic Materials **192**, 203 (1999).

8. A. E. Berkowitz and K. Takano, Journal of Magnetism and Magnetic Materials **200**, 552 (1999).

9. L. Frangou, S. Oyarzún, S. Auffret, L. Vila, S. Gambarelli, and V. Baltz, Physical Review Letters **116**, 077203 (2016).

10. W. Lin, K. Chen, S. Zhang, and C. L. Chien, Physical Review Letters **116**, 186601 (2016).

11. Z. Qiu, J. Li, D. Hou, E. Arenholz, A. T. N'Diaye, A. Tan, K.-i. Uchida, K. Sato, S. Okamoto, Y. Tserkovnyak, Z. Q. Qiu, and E. Saitoh, Nature Communications **7**, 12670 (2016).

12. H. Wang, C. Du, P. C. Hammel, and F. Yang, Physical Review Letters **113**, 097202 (2014).

13. Y. Fan, X. Ma, F. Fang, J. Zhu, Q. Li, T. P. Ma, Y. Z. Wu, Z. H. Chen, H. B. Zhao, and G. Lüpke, Physical Review B **89**, 094428 (2014).

14. W. Zhang, W. Han, S.-H. Yang, Y. Sun, Y. Zhang, B. Yan, and S. S. P. Parkin, Science Advances **2** (2016).

15. H. Chen, Q. Niu, and A. H. MacDonald, Physical Review Letters **112**, 017205 (2014).

16. D. Wu, G. Yu, C.-T. Chen, S. A. Razavi, Q. Shao, X. Li, B. Zhao, K. L. Wong, C. He, Z. Zhang, P. Khalili Amiri, and K. L. Wang, Applied Physics Letters **109**, 222401 (2016).

17. S. Woo, K. Litzius, B. Kruger, M.-Y. Im, L. Caretta, K. Richter, M. Mann, A. Krone, R. M. Reeve, M. Weigand, P. Agrawal, I. Lemesh, M.-A. Mawass, P. Fischer, M. Klaui, and G. S. D. Beach, Nat Mater **15**, 501 (2016).

18. O. Boulle, J. Vogel, H. Yang, S. Pizzini, D. de Souza Chaves, A. Locatelli, T. O. Menteş, A. Sala, L. D. Buda-Prejbeanu, O. Klein, M. Belmeguenai, Y. Roussigné, A. Stashkevich, S. M. Chérif, L. Aballe, M. Foerster, M. Chshiev, S. Auffret, I. M. Miron, and G. Gaudin, Nat Nano **11**, 449 (2016).

19. W. Jiang, P. Upadhyaya, W. Zhang, G. Yu, M. B. Jungfleisch, F. Y. Fradin, J. E. Pearson, Y. Tserkovnyak, K. L. Wang, O. Heinonen, S. G. E. te Velthuis, and A. Hoffmann, Science **349**, 283 (2015).

20. K.-S. Ryu, L. Thomas, S.-H. Yang, and S. Parkin, Nat Nano **8**, 527 (2013).

21. S. Emori, U. Bauer, S.-M. Ahn, E. Martinez, and G. S. D. Beach, Nat Mater **12**, 611 (2013).

22. G. Yu, P. Upadhyaya, X. Li, W. Li, S. K. Kim, Y. Fan, K. L. Wong, Y. Tserkovnyak, P. K. Amiri, and K. L. Wang, Nano Letters **16**, 1981 (2016).

23. J. Iwasaki, M. Mochizuki, and N. Nagaosa, Nat Nano **8**, 742 (2013).

24. A. Fert, V. Cros, and J. Sampaio, Nat Nano **8**, 152 (2013).

25. T. Schulz, R. Ritz, A. Bauer, M. Halder, M. Wagner, C. Franz, C. Pfleiderer, K. Everschor, M. Garst, and A. Rosch, Nat Phys **8**, 301 (2012).

26. O. J. Lee, L. Q. Liu, C. F. Pai, Y. Li, H. W. Tseng, P. G. Gowtham, J. P. Park, D. C. Ralph, and R. A. Buhrman, Physical Review B **89**, 024418 (2014).

27. P. P. J. Haazen, E. Murè, J. H. Franken, R. Lavrijsen, H. J. M. Swagten, and B. Koopmans, Nat Mater **12**, 299 (2013).

28. G. Yu, P. Upadhyaya, K. L. Wong, W. Jiang, J. G. Alzate, J. Tang, P. K. Amiri, and K. L. Wang, Physical Review B **89**, 104421 (2014).

29. R. E. T. S. Tacchi, M. Ahlberg, G. Gubbiotti, M. Madami, J. Åkerman, P. Landeros, arXiv:1604.02626 (2016).

30. H. T. Nembach, J. M. Shaw, M. Weiler, E. Jue, and T. J. Silva, Nat Phys **11**, 825 (2015).

31. K. Di, V. L. Zhang, H. S. Lim, S. C. Ng, M. H. Kuok, J. Yu, J. Yoon, X. Qiu, and H. Yang, Physical Review Letters **114**, 047201 (2015).

32. J. Cho, N.-H. Kim, S. Lee, J.-S. Kim, R. Lavrijsen, A. Solignac, Y. Yin, D.-S. Han, N. J. J. van Hoof, H. J. M. Swagten, B. Koopmans, and C.-Y. You, Nat Commun **6** (2015).

33. M. Belmeguenai, J.-P. Adam, Y. Roussigné, S. Eimer, T. Devolder, J.-V. Kim, S. M. Cherif, A. Stashkevich, and A. Thiaville, Physical Review B **91**, 180405 (2015).

34. K. Di, V. L. Zhang, H. S. Lim, S. C. Ng, M. H. Kuok, X. Qiu, and H. Yang, Applied Physics Letters **106**, 052403 (2015).

35. N.-H. Kim, D.-S. Han, J. Jung, J. Cho, J.-S. Kim, H. J. M. Swagten, and C.-Y. You, Applied Physics Letters **107**, 142408 (2015).

36. X. Ma, G. Yu, X. Li, T. Wang, D. Wu, K. S. Olsson, Z. Chu, K. An, J. Q. Xiao, K. L. Wang, and X. Li, Physical Review B **94**, 180408 (2016).

37. N.-H. Kim, J. Jung, J. Cho, D.-S. Han, Y. Yin, J.-S. Kim, H. J. M. Swagten, and C.-Y. You, Applied Physics Letters **108**, 142406 (2016).

38. G. Chen, T. Ma, A. T. N'Diaye, H. Kwon, C. Won, Y. Wu, and A. K. Schmid, Nature Communications **4**, 2671 (2013).

39. J. Torrejon, J. Kim, J. Sinha, S. Mitani, M. Hayashi, M. Yamanouchi, and H. Ohno, Nature Communications **5**, 4655 (2014).

40. A. Hrabec, N. A. Porter, A. Wells, M. J. Benitez, G. Burnell, S. McVitie, D. McGrouther, T. A. Moore, and C. H. Marrows, Physical Review B **90**, 020402 (2014).

41. A. Kohn, A. Kovács, R. Fan, G. J. McIntyre, R. C. C. Ward, and J. P. Goff, Scientific Reports **3**, 2412 (2013).

42. See supplementary information for detailed justifications of Eq.1, more DMI, VSM and MOKE results

43. Z. K. Wang, V. L. Zhang, H. S. Lim, S. C. Ng, M. H. Kuok, S. Jain, and A. O. Adeyeye, Applied Physics Letters **94**, 083112 (2009).

44. A. Belabbes, G. Bihlmayer, F. Bechstedt, S. Blügel, and A. Manchon, Physical Review Letters **117**, 247202 (2016).

45. J.-g. Hu, G.-j. Jin, and Y.-q. Ma, Journal of Applied Physics **94**, 2529 (2003).

46. M. T. Johnson, P. J. H. Bloemen, F. J. A. d. Broeder, and J. J. d. Vries, Reports on Progress in Physics **59**, 1409 (1996).

47. H. Yang, A. Thiaville, S. Rohart, A. Fert, and M. Chshiev, Physical Review Letters **115**, 267210 (2015).

48. W. Zhang, M. B. Jungfleisch, W. Jiang, J. E. Pearson, A. Hoffmann, F. Freimuth, and Y. Mokrousov, Physical Review Letters **113**, 196602 (2014).

49. In the polycrystalline samples investigated here, the typical AFM grain size is ~ 10 nm and a large number of grains contribute to our optical experiments with a spot size of ~ 100 um

50. M. Ali, C. H. Marrows, M. Al-Jawad, B. J. Hickey, A. Misra, U. Nowak, and K. D. Usadel, Physical Review B **68**, 214420 (2003).

51. H. Ohldag, A. Scholl, F. Nolting, E. Arenholz, S. Maat, A. T. Young, M. Carey, and J. Stöhr, Physical Review Letters **91**, 017203 (2003).